\documentclass[prl,aps,10pt,amsfonts,floatfix,superscriptaddress,longbibliography,twocolumn]{revtex4-2}

\usepackage{graphicx}
\usepackage{nicefrac}
\usepackage{amsfonts}
\usepackage{amssymb}
\usepackage{amsmath} 
\usepackage{subfigure}
\usepackage{multirow} 
\usepackage{tabularx} 
\usepackage{array}
\usepackage{float}
\usepackage{units}
\usepackage{bm}
\usepackage{hyperref}

\usepackage{xcolor}

\newcommand{\ket}[1]{\left| #1 \right>} % for Dirac bras
\newcommand{\bra}[1]{\left< #1 \right|} % for Dirac kets
\begin{document}
\title{Effects of auto-correlated disorder on the dynamics in the vicinity of the many-body localization transition}
\author{Isaías Vallejo-Fabila}
\author{E. Jonathan Torres-Herrera}
\affiliation{Instituto de F\'isica, Benem\'erita Universidad Aut\'onoma de Puebla,
Apt. Postal J-48, Puebla, 72570, Mexico}

\begin{abstract}
The presence of frozen uncorrelated random on-site potential in interacting quantum systems can induce a transition from an ergodic phase to a localized one, the so-called many-body localization. Here we numerically study the effects of auto-correlated disorder on the static and dynamical properties of a one-dimensional many-body quantum system which exhibits many-body localization. Specifically, by means of some standard measures of energy level repulsion and localization of energy eigenstates, we show that a strong degree of correlations between the on-site potentials in the one-dimensional spin-1/2 Heisenberg model leads to suppression of the many-body localization phase, while level repulsion is mitigated for small disorder strengths, although energy eigenstates remain well extended. Our findings are also remarkably manifested in time domain, on which we put main emphasis, as shown by the time evolution of experimentally relevant observables, like the return probability and the spin auto-correlation function.
\end{abstract}

\maketitle
%%%%%%%%%%%%%%%%% Introduction %%%%%%%%%%%%%%%%%%%%
The celebrated, very well studied and equally understood single-particle Anderson localization (AL)~\cite{Anderson1958} is expanded in some sense to the realm of interacting quantum systems by the still incompletely understood so-called many-body localization (MBL), where a thermal description of the system properties fails~\cite{Nandkishore2015,Alet2018,Altman2018,Abanin2019}. Among some questions challenging MBL and some of its features are the ones related to the role of finite-size effects~\cite{Potirniche2019,abanin2021,Sierant2022}, finite times~\cite{vsuntajs2020quantum} and self-averaging~\cite{Schiulaz2020,TorresHerrera2020a,TorresHerrera2020b}. Deepen our knowledge about MBL is of fundamental theoretical, experimental~\cite{Schreiber2015,Bordia2016,Kohlert2019} and technological [see for instance~\cite{Halpern2019}] interest.

The standard scenario in the non-interacting case is that AL transition occurs in three dimensions for strong enough disorder strength, meanwhile in one- and two-dimensional systems at the limit of infinite system size an infinitesimal strength of the uncorrelated disorder is enough to localize all eigenstates of the system (see~\cite{abrahams1979,Lee1985,Lagendijk2009} and references therein). However, it has been shown, theoretical and experimentally, that particular correlations between the on-site potentials are able to produce delocalization and even the appearance of a mobility edge in one-dimensional systems~\cite{dunlap1990,deMoura1998,izrailev1999,kuhl2000,kuhl2008} (see also~\cite{izrailev2012} and references therein). Further studies of auto-correlated disorder in low-dimensional non-interacting systems have addressed, for instance, the onset of Bloch-like oscillations ~\cite{dominguez2003},  localization in ultracold atoms~\cite{sanchez2007,lugan2009} and optical lattices~\cite{fratini2015}, violation of Harris criterion~\cite{shima2004}, non-universality of AL~\cite{titov2005} and features of the entanglement spectrum~\cite{mondragon2013char,pouranvari2014,andrade2014}. Recently, the effects of auto-correlated disorder have been also investigated in the joint context of ultracold atoms and machine learning~\cite{pilati2019}. A natural question is what could be the effects of auto-correlated disorder in the context of MBL, however, in contrast with the non-interacting case, effects of auto-correlated disorder in interacting many-body quantum systems have been less explored~\cite{Dukesz2009,xu2019,ghosh2019,roy2020,samanta2021,darius2022}. While most works are focused on time-independent quantities, dynamics are predominantly left aside even though they are of fundamental relevance not only by their ubiquity but from a theoretical point of view and also for experiments dealing with many-body quantum systems, like the ones with ultracold atoms in optical lattices~\cite{Bloch2008,Schreiber2015,bernien2017,ebadi2021}, trapped ions~\cite{leibfried2003} and superconducting qubit arrays~\cite{gong2021}, where information about the system properties is usually accessed through the dynamics. It is fair to mention that in~\cite{maksymov2020} the role of speckle disorder on the dynamics of the imbalance was part of the analysis that led to the main conclusion that MBL transitions under uncorrelated disorder and speckle disorder belong to the same universality class. Here, however, our aim is not to determine the universality class of MBL or even the behavior of critical points under our model of auto-correlated disorder, but instead to provide a general description of its effects on the whole time evolution of the many-body quantum system that we study around the MBL transition. 

%%%%%%%%%%%%%%%%% Model %%%%%%%%%%%%%%%%%%%%
For our purposes we consider the Heisenberg model for spin-1/2 particles in a one-dimensional lattice with pair interactions, on-site potentials and periodic boundaries, 
\begin{equation}\label{eq:Ham}
H=\sum_{k=1}^{L} ( S_{k}^{x}S_{k+1}^{x}+S_{k}^{y}S_{k+1}^{y} +S_{k}^{z}S_{k+1}^{z} ) + \sum_{k=1}^{L}h_{k}S_{k}^{z}.  
\end{equation}
In Eq.~\eqref{eq:Ham} we have set $\hbar=1$ and $S_{k}^{x,y,z}=\sigma_{k}^{x,y,z}/2$ are spin-1/2 operators in terms of Pauli matrices acting over the particle located at site $k$. Traditional studies consider the amplitudes $h_k$ being uncorrelated random numbers from an uniform distribution, that is, $h_k\in U(-h,h)$, with $h$ the disorder strength. The critical point $h_c$ for MBL is not exactly known but there exist some estimates for finite-size systems, from~\cite{Oganesyan2007,Pal2010,Berkelbach2010,Kjall2014,Luitz2015} the  bounds are $3< h_c<4$, meanwhile in~\cite{Devakul2015,Doggen2018,Sierant_2020a,Sierant_2020} an upper bound $h_c>4$ is given. Here we consider the amplitudes $h_k$ also as random numbers from an uniform distribution with support $[-h,h]$, but linearly correlated, that is, $\mathbb{E}(h_k h_{\ell})\neq \mathbb{E}(h_k)\mathbb{E}(h_{\ell})$ for $k,\ell=1,2,\dots L$ and $\mathbb{E}$ stands for the statistical expectation value.
The method that we use to generate auto-correlated random numbers is based on the cumulative distribution function $F(X)$ of the sum of a $U(0,1)$ number and a $U(0,1/c)$ number, where $c$ is any positive real number. With $V_{k}$ an \textit{i.i.d.} $U(0,1)$ we generate the sequence 
\begin{equation}
  \begin{aligned}
X_{1} &= V_{0} + V_{1}/c,\\ 
X_{k} &= F(X_{k-1}) + V_{k}/c, \hspace{0.5cm} k>1.
\end{aligned}
\end{equation}
such that the sequence of $F(X_{k})$ is auto-correlated $U(0,1)$, then the random numbers $h_k$ in Hamiltonian~\eqref{eq:Ham} are obtained according to $h_k=h[2F(X_{k})-1]$. The distribution $F(X_{k})$ depends on the value of $c$ and it can be determined exactly as shown in~\cite{Thomas1993}. By tuning $c$ any desired degree of correlations is obtained, $c\approx0$ means absence of correlations, meanwhile the strength of correlations grow larger as $c$ increases. 

Since Hamiltonian~\eqref{eq:Ham} preserves the total magnetization in $z$-direction, ${\cal{S}}^z=\sum_k S_k^z$, the Hilbert space splits in sectors with fixed values of ${\cal{S}}^z$. Here we work in the subsector with ${\cal{S}}^z=0$ for which the dimension is ${\cal{D}}=L![(L/2)!]^{-2}$.

\paragraph{Quantities.-}\par
We describe the static quantities, as well as the time-dependent ones that will be used in our analysis as indicators of ergodicity and localization.  

We start by denoting the eigenvalues and eigenstates of Hamiltonian~\eqref{eq:Ham} by $E_{\alpha}$ and $\ket{\psi_{\alpha}}$ respectively. Next, to analyze level repulsion we use the mean value of the \emph{ratio between consecutive level spacings} $\overline{r}$~\cite{Oganesyan2007}, defined through
\begin{equation}\label{eq:r}
 \overline{r}= \frac{1}{{\cal{D}}}\sum_{\alpha=1}^{\cal{D}}\text{min}\left(r_{\alpha},\frac{1}{r_{\alpha}}\right),\,\,\text{with}\,\, r_{\alpha}=\frac{E_{\alpha+1}-E_{\alpha}}{E_{\alpha}-E_{\alpha-1}}.  
\end{equation}
Level repulsion is absent in the MBL phase, corresponding to spacings with Poisson-like statistics and $\overline{r}\approx 0.386$~\cite{Oganesyan2007,Atas2013}. Meanwhile, in the ergodic phase the spacings have GOE-like statistics, that is, energy eigenvalues present linear repulsion, for which $\overline{r}\approx 0.536$~\cite{Atas2013}.  The ratio decreases monotonically between those two limits as the magnitude of the uncorrelated disorder increases~\cite{Oganesyan2007}.

To study the degree of localization of the eigenstates of Hamiltonian~\eqref{eq:Ham} we use the
\emph{inverse participation ratio}, IPR, given by 
\begin{equation}\label{eq:IPR}
\text{IPR}_\alpha=\sum_{n=1}^{\cal{D}}\left|C_{n}^{\alpha}\right|^4,\,\,\text{with}\,\,C_{n}^{\alpha}=\left\langle\psi_{\alpha}|n\right\rangle.
\end{equation}
Where $\left\{\ket{n}\right\}$ represents a suitable basis determined by physical considerations. Since our interest is localization in real space, we choose the basis composed by eigenstates of the $z$-part of Hamiltonian~\eqref{eq:Ham}, so-called site basis. A state is extended in the respective basis if $\text{IPR}_{\alpha}\propto {\cal{D}}^{-1}$ [for instance, $\text{IPR}_\alpha=3/({\cal{D}}+2)$ for eigenstates of GOE matrices~\cite{Ullah1964}], while for localized states $\text{IPR}_{\alpha}$ is ${\cal{O}}(1)$.

At a first glance it could appear as redundant to analyze both quantities, but up to our knowledge no one-to-one correspondence between the behavior of $\overline{r}$ and $IPR_{\alpha}$ for model~\eqref{eq:Ham} has been proved. Certainly a proof of this last point is not the aim of this work, however, the analysis of both features, level repulsion and degree of localization of states, is important for our objectives and because they show up in different stages of the time evolution of the system~\cite{Torres2014NJP,Torres2015,Torres2017Philo,Torres2018}.

Dynamics can be analyzed employing the \emph{return probability}, which is also known as survival probability and non-decay probability. It measures the time dependent probability for a given quantum system to return to its initial state $|\left.\Psi(0)\right\rangle$, it is given by
\begin{equation}\label{eq:SP}
 RP(t)=\left|\left\langle\Psi(0)|e^{-iHt}|\Psi(0)\right\rangle\right|^2.%=\left|\sum_{\alpha=1}^{\cal{D}}|C_{\alpha}^{0}|^2e^{-iE_{\alpha}t}\right|^2.
\end{equation}
Return probability has been analyzed in several contexts, both theoretical and experimentally. By the side of experiments in molecules~\cite{rothe2006violation}, ultra-cold atoms in magneto-optic traps~\cite{wilkinson1997}, atom chip~\cite{gherardini2017} and prethermalization in Floquet systems~\cite{singh2019}. See also an interesting proposal in~\cite{tikhonov2021} regarding the experimental measuring of $RP(t)$. Theoretically in relation with the time evolution of unstable quantum states~\cite{chiu1977time} and possible implications in cosmology of the late time behavior of false vacuum decay~\cite{krauss2008late}, non-periodic substitution potentials~\cite{de1999quantum}, non-inteacting fermions~\cite{krapivsky2018quantum}, quantum walks and complex networks~\cite{ampadu2012return,mulken2011,xu2008,riascos2015}, matter-radiation interaction models like Dicke ~\cite{lerma2019} and Bose-Hubbard~\cite{delacruz2020}.

In addition, to further study dynamics we consider the \emph{spin auto-correlation function},
\begin{equation}\label{eq:SCF}
I(t)=\frac{4}{L}\sum_{k=1}^{L}\bra{\Psi(0)}S_{k}^{z}(0)S_{k}^{z}(t)\ket{\Psi(0)}.
\end{equation}
This quantity measures the similarity between the initial spin configuration and the one at a late time $t$. For specific initial states, such as a N{\'e}el-like state,  it coincides with the density imbalance measured in experiments with cold atoms~\cite{Schreiber2015}.

Since we are dealing with a disordered system, we compute averages over disorder realizations for the static quantities, $\overline{r}$ and $\text{IPR}_{\alpha}$, while an additional average over initial states is done for the dynamical quantities, $RP(t)$ and $I(t)$. Our results will be presented using the notation for the averages as $\left\langle\dots\right\rangle$ irrespective of their kind. The initial states we use are part of the set $\left\{\ket{n}\right\}$, that is, $\ket{\Psi(0)}\equiv \ket{n}$ with energy $E_n=\bra{n}H\ket{n}\approx 0$.

Before presenting our results, we judge as instructive and pertinent to give an overview of the existing picture about the dynamics of many-body quantum systems with uncorrelated disorder. Certainly, the dynamics of many-body quantum systems could be approached by many kinds of observables, but for decaying-in-time observables, like the return probability and the spin auto-correlation function, a clear picture already exists~\cite{Torres2018}. For the first one, quantum mechanics predicts a universal quadratic decay in time for very short times, this is the region of the ancient Zeno's paradox in the quantum realm ~\cite{misra1977zeno,pascazio2014all}. The quadratic decay is followed by one which depends on the interaction strength, exponential for weak strengths and Gaussian for strong enough strengths~\cite{Torres2014NJP}. A power-law decay is developed at larger times, it can have two different sources, one being the inevitable bounds in the energy spectrum~\cite{Khalfin1958} and the other one related to the increasing of correlations between eigenstates components as the disorder strength increases~\cite{Torres2015}. The former one has been used to anticipate thermalization~\cite{Tavora2016,Tavora2017} in interacting quantum systems, while the last one was investigated in relation with the multifractality of eigenstates~\cite{Torres2015,Torres2017}. Power-law decays have been also observed for the imbalance, both theoretically~\cite{luitz2016,Doggen2018} and experimentally~\cite{Luschen2017}. Eigenstates correlations and their relation with the return probability was also studied more recently in~\cite{tikhonov2021}. At even larger times and preceding saturation the repulsion hole appears, a dynamical manifestation of short- and long-range energy-level repulsion which explicitly occurs at the many-body Thouless time and ends up at the longest time scale, the so-called Heisenberg time. In particular, it was shown that the ratio between the Thouless and Heisenberg times approaches unity as the system approaches the MBL phase~\cite{Schiulaz2019}, this fact was further confirmed later in~\cite{vsuntajs2020quantum}. 

\paragraph{Results.-}\par
\begin{figure}
    \centering
    \includegraphics[scale=0.385]{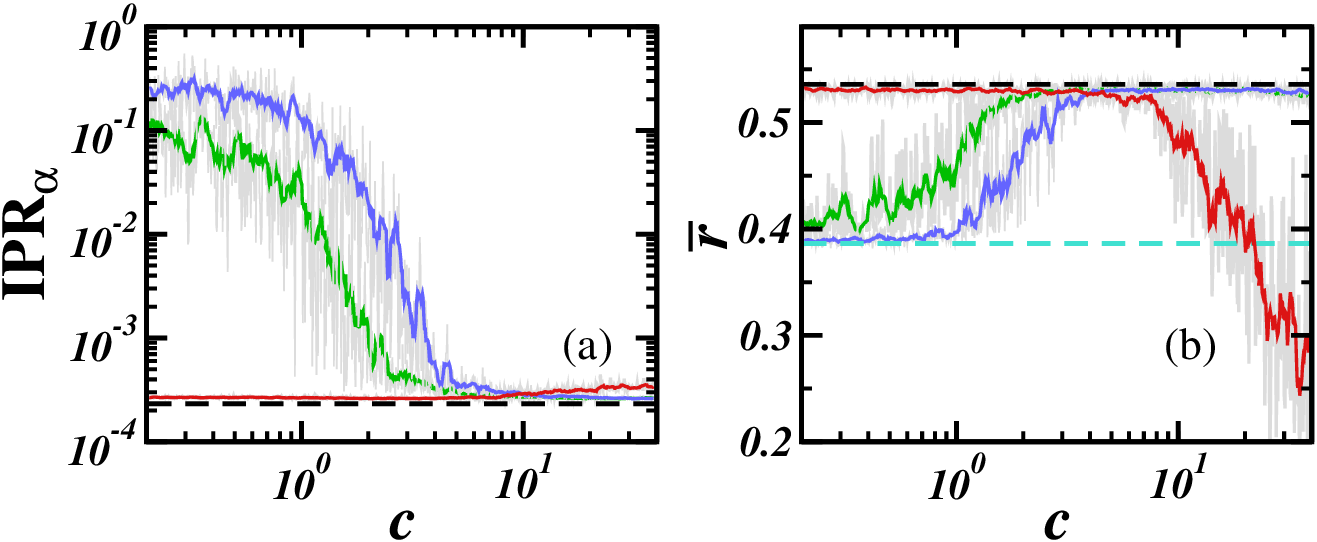}
    \caption{Inverse participation ratio (IPR) and average spacing ratio $\overline{r}$ versus degree of correlations $c$. (a) IPR from Eq.~\ref{eq:IPR} and (b) $\overline{r}$ from 
    Eq.~\ref{eq:r}. In each panel the shaded (gray) curves represent raw data from a single disorder realization (see main text), while the curves in red, green and blue are running averages for $h=0.5$, $h=3.75$ and $h=6$, respectively. Theoretical values for GOE (black dashed line) and Poisson (turquoise dashed line). System size is $L=16$.}
    \label{fig:IPR_r}
\end{figure}

 We now present our results, explicitly we show the effects of the correlations between the on-site random potentials on the static and time-dependent properties previously defined. We start with the behavior of the structure of eigenstates [Fig.~\ref{fig:IPR_r}(a)] and repulsion between adjacent energy levels [Fig.~\ref{fig:IPR_r}(b)] with respect to the degree of correlations represented by $c$ for a fixed system size $L=16$. In both panels the results are for three different disorder strengths, $h=0.5$ (red solid curve), $h=3.75$ (green solid curve) and $h=6.0$ (blue solid curve), which in a context of uncorrelated disorder are representative of the ergodic, critical and localized regions, respectively. We use dashed lines to represent the GOE theoretical prediction (black) and Poisson prediction (turquoise) for both quantities, $\text{IPR}_{\alpha}$ and $\overline{r}$. We observe two different behaviors depending on the disorder strength as the degree of correlations increases from a very small value of $c$ to a larger one, $c\approx 20 $. For small disorder strength, $h = 0.5$, the $\text{IPR}_\alpha$ almost does not change with respect to $c$. However, level statistics clearly transit from GOE-like to different statistics, that approach and eventually cross down Poisson-like statistics as $c$ increases, reaching a regime of the Schnirelman peak (see~\cite{SMM} for some specific examples of the level spacing distribution and see, also, Refs.~\cite{Shnirelman1975,Chirikov1995,Frahm1997,Chakrabarti2002,Zangara2013} therein). For the disorder strengths, $h=3.75$ and $h=6.0$, the behaviors are similar, a crossover occurs as $c$ increases from localized states to extended states and from Poisson-like statistics to GOE-like. It is worth to mention that results from a single disorder realization of the static properties show big fluctuations, this is depicted by the shaded (gray) curves in both panels of Fig.~\ref{fig:IPR_r}. 
Of course, those fluctuations can be reduced if an average over disorder realizations is carried out, see~\cite{SMM} for further details. Also note that in Fig.~\ref{fig:IPR_r}(a) an average over $100$ energy eigenstates with energy closest to zero was performed, however this was not enough to smooth the curves for $h=3.75$ and $h=6.0$ because $\text{IPR}_{\alpha}$ is not self-averaging for those values~\cite{solorzano2021}. To circumvent this situation, we decided to present in a simple way running averages over intervals of $10$ values of the static properties along $c$ which are precisely represented by the colored solid lines already explained above.

\begin{figure}
    \centering
    \includegraphics[scale=0.42]{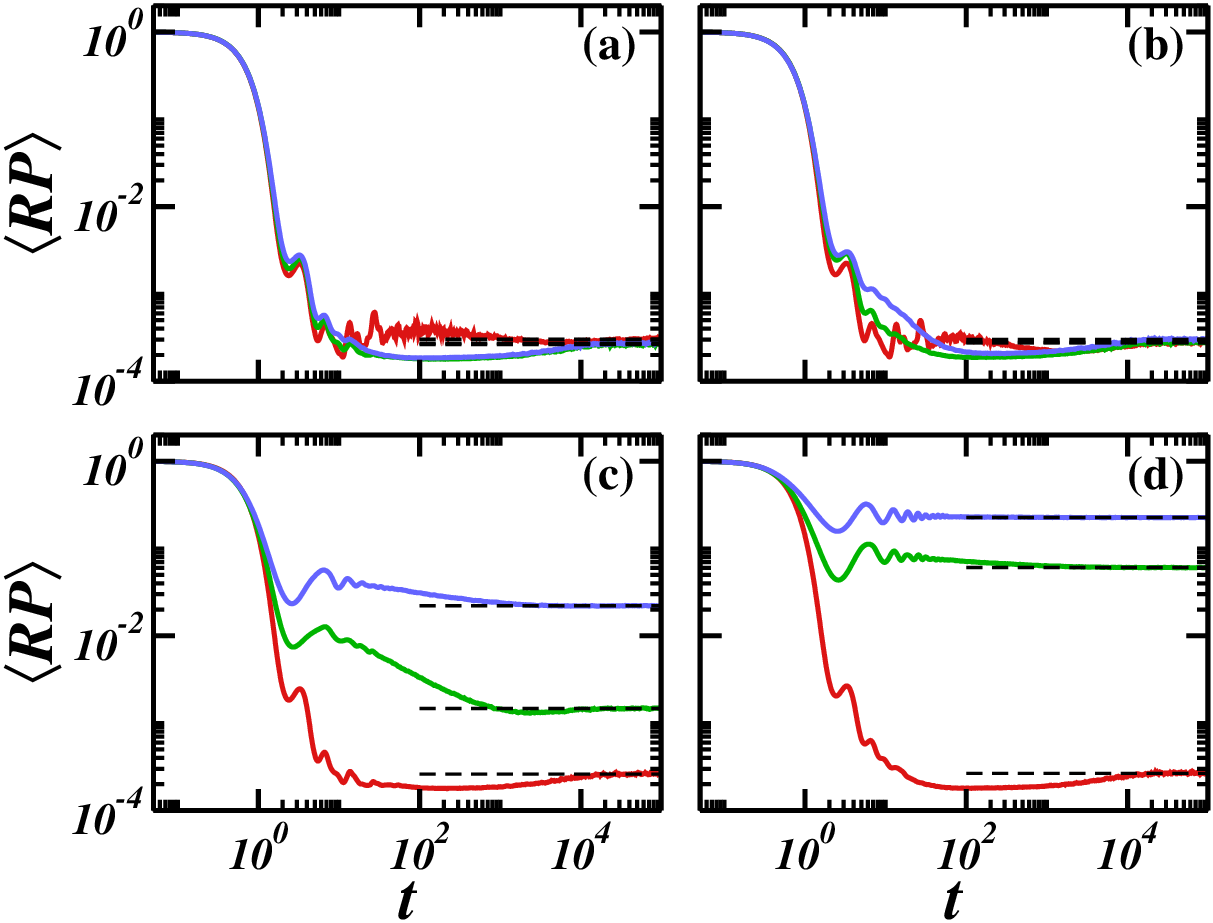}
    \caption{Averaged return probability for different disorder strengths and degrees of correlations $c$. (a) $c=15$, (b) $c=8$, (c) $c=2$, (d) $c=0.5$. In each panel $h=0.5$ (red), $h=3.75$ (green), $h=6$ (blue). Dashed lines are the corresponding saturation values. System size is $L=16$. An average over $25\times10^3$ samples ($500$ initial states and $50$ disorder realizations). An additional running average was implemented in order to smooth even more the curves.}
    \label{fig:SP}
\end{figure}

We now move to the time-dependent quantities, in Fig.~\ref{fig:SP} we depict the behavior of the return probability for three different disorder strengths, just like in Fig.~\ref{fig:IPR_r} and with the same coloring code. Four values of $c$ are selected from large to small in order to cover different regimes of eigestates structure and level statistics, namely $c=15$ (a), $c=8$ (b), $c=2$ (c) and $c=0.5$ (d) (see Fig.~\ref{fig:IPR_r}). 
The system size is also fixed to $L=16$. In general we observe similar initial dynamics for the four values of $c$ and the three values of $h$, that is, a typical initial non-exponential decay, explicitly a Gaussian decay~\cite{izrailev2006,Torres2014PRA,Torres2014PRAb,Torres2014NJP} which is followed by some fluctuations related to the details of the local density of states which is not a perfect Gaussian~\cite{Torres2014PRA,Torres2014PRAb,Torres2014NJP,Torres2015}. For small disorder, $h=0.5$ (red solid line), the power-law decay with exponent $2$ predicted for weak uncorrelated disorder~\cite{Tavora2016,Tavora2017} is observed in the approximated time interval $t\in[3,12]$ for all values of $c$. In the limit of large system sizes the power-law decay with exponent 2 prevails for longer times~\cite{torres2019}. The survival of power-law decays with system size was also studied for the imbalance in~\cite{Doggen2018}. The repulsion hole is wiped out for $c=15$, meanwhile the hole appears and becomes more clear as the value of $c$ decreases, as shown for $c=8$ and $c=2$ in Figs.~\ref{fig:SP}(b) and (c) respectively. The saturation values of the return probability which corresponds to the average IPR of the initial state in the energy basis and indicated by corresponding dashed lines in Fig.~\ref{fig:SP} remain almost the same and with a small value, thus indicating that the average IPR of the initial states behaves at least in a similar way as the energy eigenstates [see Fig.~\ref{fig:IPR_r}(a)]. For the two other disorder strengths, $h=3.75$ (green solid curves) and $h=6.0$ (blue solid curves), the dependence on $c$ of the late-time dynamics of the return probability is quite different from the case $h=0.5$. In particular, for $c=15$ and both values of $h$ there is practically no distinction between the corresponding dynamics, they show a similar repulsion hole and saturation value. The curves for the two different values of $h$ get more separated as the value of $c$ increases. The behaviors of the repulsion hole and the saturation value are consistent with the results of the static quantities previously presented, while the dependence of the power-law decay on $c$ and $h$ deserves further studies. For $c=0.5$ in Fig.~\ref{fig:SP}(d) the degree of correlations is weak enough to recover the known picture for uncorrelated disorder~\cite{Torres2015}. 
\begin{figure}[!ht]
    \centering
    \includegraphics[scale=0.42]{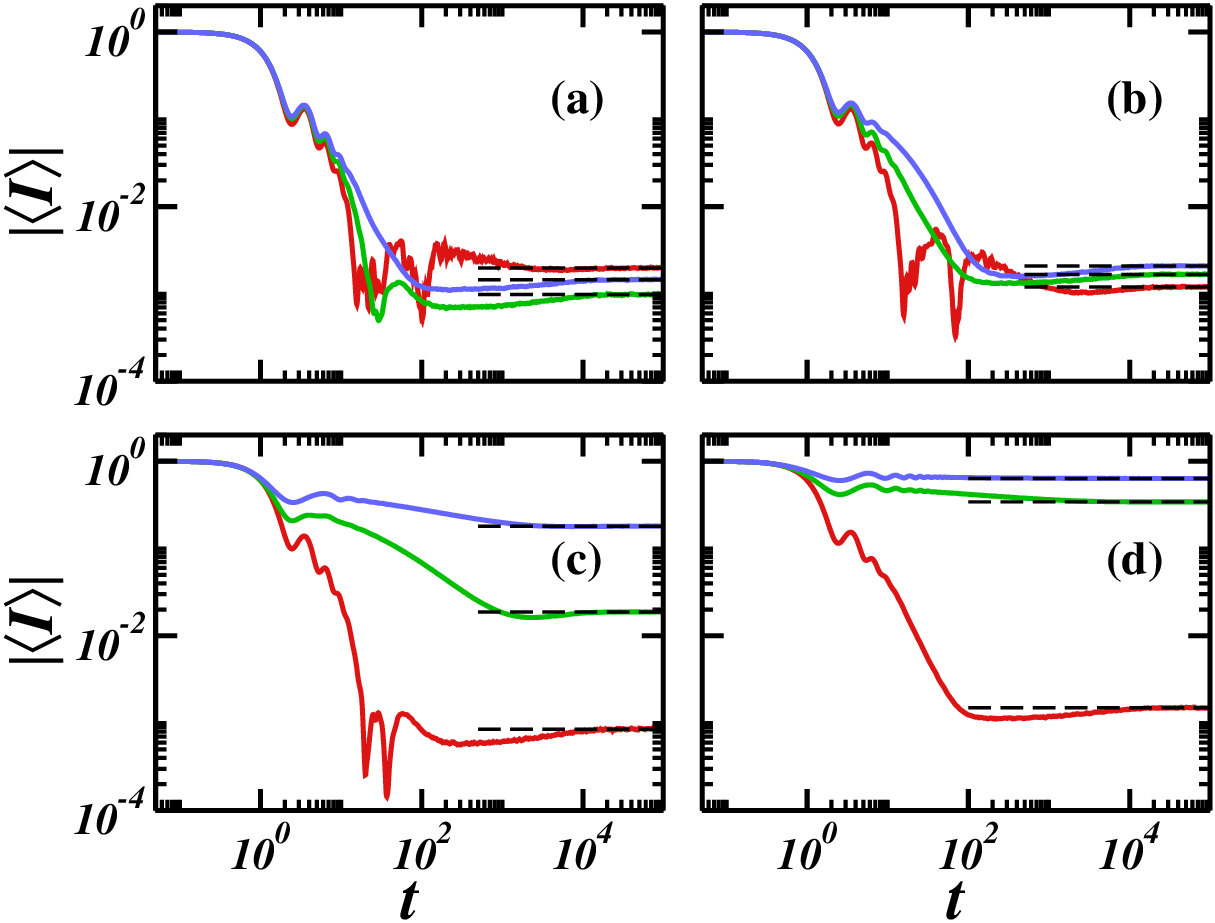}
    \caption{Absolute value of the averaged spin auto-correlation function for same parameters, coloring code and averaging procedure as used in Fig.~\ref{fig:SP}. Dashed lines are the corresponding saturation values of $|\left\langle I(t)\right\rangle|$.}
    \label{fig:Imb}
\end{figure}
We finish our analysis by considering the time evolution of the spin auto-correlation function, $I(t)$ [Eq.~\eqref{eq:SCF}]. Since $I(t)$ can have positive and negative values even after an average over initial states and disorder realizations, in Fig.~\ref{fig:Imb} we present our results for the absolute value of the averaged $I(t)$. The same parameters as for $RP(t)$ in Fig.~\ref{fig:SP} were used. Once again we observe that the inclusion of correlations between the on-site random potentials leads to rich and nontrivial dynamics, but now testified by $I(t)$. The dependence on $c$ of this quantity is similar to that of $RP(t)$. For large $c$ [Fig~\ref{fig:Imb}(a)] the evolution for $h=3.75$ and $h=6.0$ shows a repulsion hole and small saturation values, meanwhile for $h=0.5$ there is no hole. At this point we should note that the negative values of the averaged $I(t)$ are present only for $h=0.5$ and $c=15,8,2$ as seen by the bumps inside the time interval $t\in[10,100]$ of Figs.~\ref{fig:Imb}(a), (b) and (c). The behaviors of $I(t)$ for all disorder strengths, when the value of $c$ decreases, approaches the ones for uncorrelated disorder~\cite{luitz2016,Torres2018}. Although the repulsion hole is present in $RP(t)$ and $I(t)$, it should be noted that in~\cite{lezama2021} a careful analysis was performed to show that while for $RP(t)$ the depth of the repulsion hole increases, for $I(t)$ and other observables the depth apparently decreases when the system size increases. The persistence of the observed effects due to correlated disorder in the $L\to\infty$ limit is also another natural and interesting question. We address this question in~\cite{SMM}. 

\paragraph{Conclusions.-}\par
We revealed that the inclusion of linearly auto-correlated on-site random potentials in a paradigmatic model for many-body localization and tuning its strength lets to control both, static and dynamical properties of the system in a more rich and nontrivial way than uncorrelated disorder. Strong correlations result in thermal-like behavior of the studied dynamical probes when the disorder strength is large. Meanwhile for small disorder strength only energy-level indicators are significantly affected. Diminishing the degree of correlations leads to the usual behavior when uncorrelated disorder is considered. We hope that our results will motivate the community to consider different observables of interest in theoretical and experimental studies. Additional studies could also address questions related to time scales and power-law decays. 

%%%%%%%%%%%%%%%%%%%% ACKNOWLEDGMENTS %%%%%%%%%%%%%%%%%%%%%
\begin{acknowledgments}
I.V.-F. and E.J.T.-H. are grateful to LNS-BUAP for their supercomputing facility. E.J.T.-H. acknowledges financial support from VIEP-BUAP, project No. 00270.
\end{acknowledgments}


%\bibliography{Correlated.bib}

\end{document}

\title{Supplemental Material:\\
Effects of auto-correlated disorder on the dynamics in the vicinity of the many-body localization transition}

\author{Isaías Vallejo-Fabila}
\author{E. Jonathan Torres-Herrera}
\affiliation{Instituto de F\'isica, Benem\'erita Universidad Aut\'onoma de Puebla,
Apt. Postal J-48, Puebla, 72570, Mexico}

\maketitle

In this supplemental material (SM), we analyze the dependence on system size $L$ of the time-independent quantities described in the main text. Those include the averaged ratio between nearest-neighbor energy levels spacings $\left\langle\tilde{r}\right\rangle$, level spacing distribution $P(s)$ and the inverse participation ratio of energy eigenstates, $\text{IPR}_\alpha$. The analysis made in this SM provides a more solid ground for our claims done in the main text.

\section{Level statistics}
\begin{figure}[h]
\begin{center}
\includegraphics[width=8.65cm]{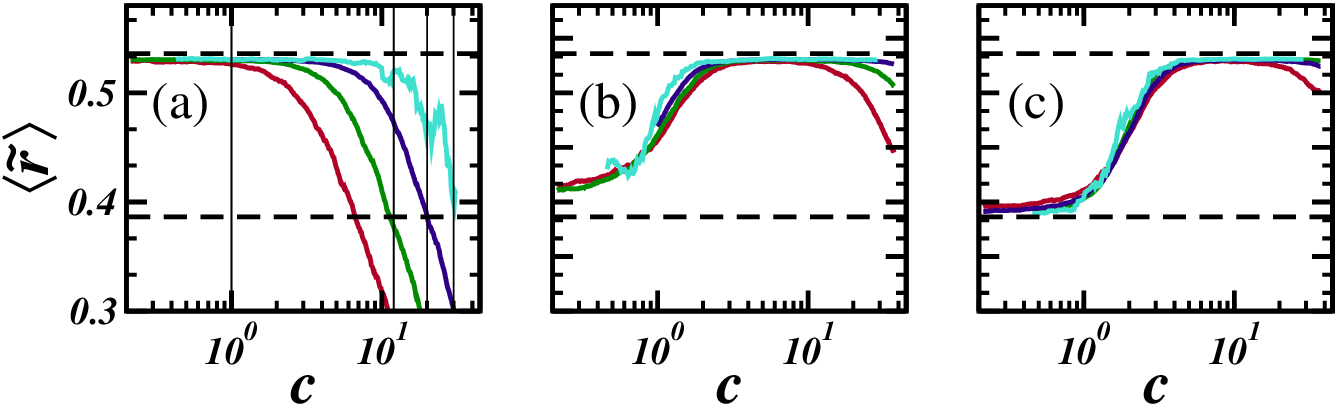}
\caption{Averaged level spacings ratio versus degree of correlations $c$, for different disorder strengths and system sizes. (a) $h=0.5$, (b) $h=3.75$ and (c) $h=6.0$. System sizes are $L=12$ (red), $L=14$ (green), $L=16$ (blue) and $L=18$ (turquiose). Theoretical values for GOE (black upper line) and Poisson (black lower line). The vertical solid lines mark points used in Fig.~\ref{fig:SM2} for $L=16$.}
\label{fig:SM1}
\end{center}
\end{figure}

In Fig.~\ref{fig:SM1} we show $\left\langle\tilde{r}\right\rangle$ [see Eq.~(3) of the main text] for three different disorder strengths and four different even system sizes, from $L=12$ to $L=18$. When the disorder strength is small, $h=0.5$ [Fig.~\ref{fig:SM1}(a)], and for any fixed size, level statistics transit from GOE-like (dashed black horizontal line) for small degree of correlations ($c$ small) to different statistics, eventually passing through Poisson-like (dashed magenta horizontal line) but going down even further for strong degree of correlations ($c$ large). However, the value of $c$ for which $\left\langle\tilde{r}\right\rangle$ leaves the GOE value  increases as the system size increases. Our results for the four systems sizes could indicate that in the $L\to\infty$ limit the system is insensitive to correlations between disorder on-site potentials, but this point deserves further efforts.  
%==============Figure1===========
%=================================
Panels (b) and (c) of Fig.~\ref{fig:SM1} shows results for $h=3.75$ and $h=6.0$ respectively. There the picture is contrasting with the one for $h=0.5$, not only in the behavior already described around Fig.~1 of the main text but also in the system size dependence. We observe that for small degree of correlations, $c\lesssim 10$, the dependence on $L$ looks weak for $h=3.75$ and surprisingly it is absent for $h=6.0$, since in the further case the curves for different system sizes almost superimpose while in the last one the curves superimpose. For strong degree of correlations and both values of disorder strength $h=3.75$ and $h=6.0$,  the curves approach the GOE value as the system size increases, indicating that it could happen that in the limit of a large system size enough to completely wipe out finite-size effects, level statistics remain GOE-like. 

Note that in order to count with even more confidence in the trends dictated by our results, we obtained the curves depicted in Figs.~\ref{fig:SM1} and~\ref{fig:SM3} by performing an average over $8\times 10^3$ samples obtained with $200$ energy eigenstates from each of $40$ disorder realizations for $L$ from $12$ up to $16$. For $L=18$ a total of $200$ samples obtained with $200$ energy eigenstates from a single disorder realization were used for the average. An additional running average was performed to reduce even more the fluctuations.  

%============Figure2============
\begin{figure}[htp]
\begin{center}
{\includegraphics[width=6.5cm]{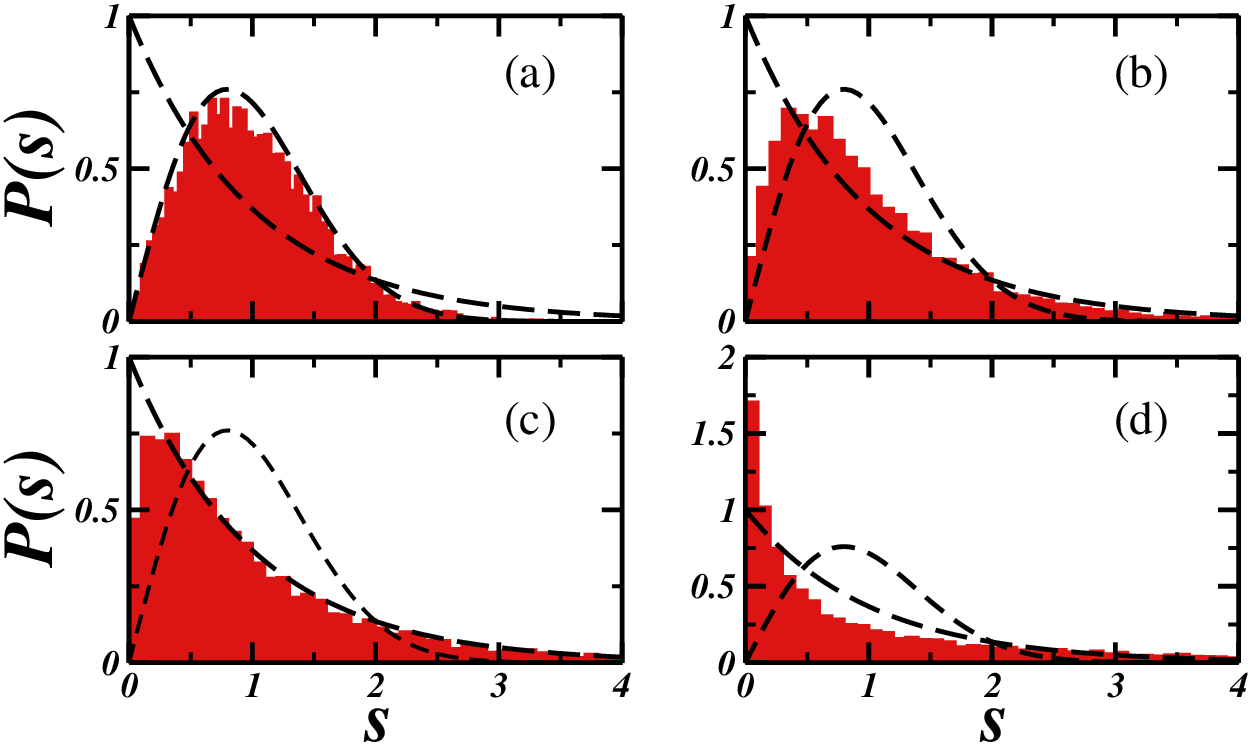}}
\caption{Level spacing distribution, $P(s)$, for $h=0.5$ and different values of $c$, see Fig.~\ref{fig:SM1}. (a) $c=1$, (b) $c=12$, (c) $c=20$ and (d) $c=30$. Dashed curves are theoretical predictions for Poisson and GOE $P(s)$. System size is $L=16$ and a single realization is considered.}
\label{fig:SM2}
\end{center}
\end{figure} 
%==============================

Figure~\ref{fig:SM2} depicts some examples of the level spacing distribution $P(s)$ for the case $h=0.5$, fixed system size $L=16$ and four degrees of correlations as described in the caption and corresponding to values of $c$ marked by the vertical lines in Fig.~\ref{fig:SM1}. It is known that in the absence of level repulsion $P(s)=\exp(-s)$, meanwhile for GOE-like level repulsion the distribution is $P(s)=\frac{\pi}{2}s\exp(-\frac{\pi}{4}s^2)$. For panel (a) with $c=1$ the distribution is GOE-like, for (b) with $c=12$ the distribution has an intermediate shape between GOE and Poisson, arriving to $c=20$ the shape is Poisson-like and finally, in (d) with $c=30$ the distribution presents a large peak at $s=0$, which is a signature of symmetries and corresponding degeneracies in the energy spectrum~\cite{Shnirelman1975,Chirikov1995}. The so-called Schnirelman peak has been observed also in rough billiards~\cite{Frahm1997}, Calogero-like three-body problem~\cite{Chakrabarti2002} and more recently in one-dimensional spin-1/2 models~\cite{Zangara2013}.

\section{Structure of energy eigenstates}
We now turn our attention to the dependence on system size of the inverse participation ratio, $\text{IPR}_\alpha$, as defined by Eq.~(4) of the main text. 

\begin{figure}[htp]
\begin{center}
{\includegraphics[width=8.5cm]{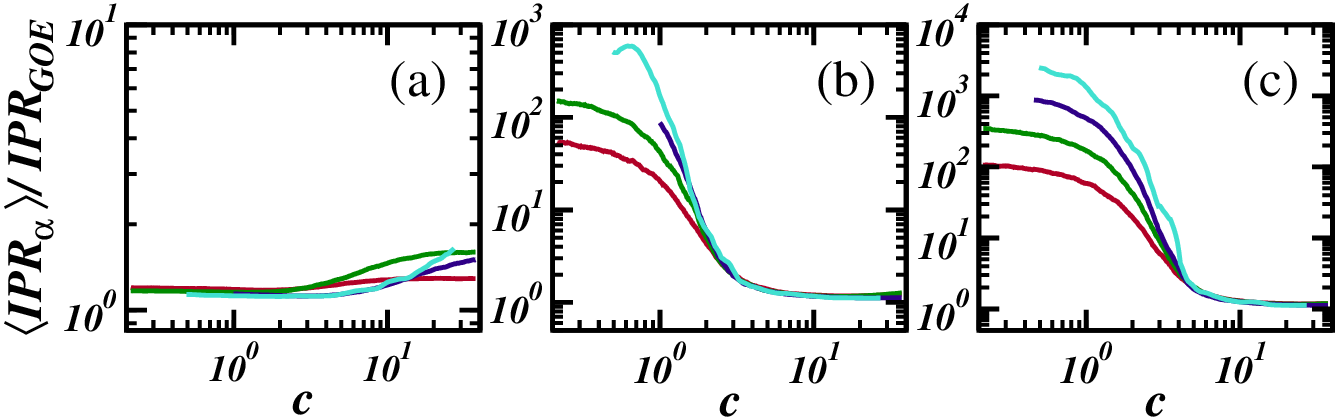}}
\caption{Normalized inverse participation ratio vs degree of correlations $c$. Same values of disorder strengths, colors for system sizes and number of samples as in Fig.~\ref{fig:SM1}.}
\label{fig:SM3}
\end{center}
\end{figure}

We present in Fig.~\ref{fig:SM3} the inverse participation ratio of energy eigenstates of Hamiltonian~(1) versus the degree of correlations $c$ and for the same system sizes as in Fig.~\ref{fig:SM1}. $\text{IPR}_\alpha$ is shown normalized with the theoretical prediction for GOE, $\text{IPR}_\text{GOE}=3/({\cal{D}}+2)$. Panel (a) depicts results for $h=0.5$, we observe that eigenstates remain well extended like GOE eigestates for all values of $c$ and all considered system sizes, since the ratio $\left\langle\text{IPR}_\alpha\right\rangle/\text{IPR}_\text{GOE}$ stays close to 1. The dependence on system size for small disorder strength contrasts with the behavior of $\left\langle\tilde{r}\right\rangle$ [see Fig.~\ref{fig:SM1} (a)].  
%============Figure3============
%==============================

Figs.~\ref{fig:SM3} (b) and (c) show a similar trend for both disorder strengths, $h=3.75$ and $h=6.0$. As anticipated in the main text, energy eigenstates are localized for fixed weak correlations, but certainly they are more localized for $h=3.75$ than for $h=6.0$. The ratio $\left\langle\text{IPR}_\alpha\right\rangle/\text{IPR}_\text{GOE}$ gets much larger than 1 and its value increases with system size. Note that for $L=16$ and $L=18$ our data for small $c$ are incomplete, but the tendency is clear enough. The level of localization decreases as $c$ increases, but the energy eigenstates get maximally extended when $c\approx 4$ for $h=3.75$ and $c\approx 5$ for $h=6.0$, from this point there is no system size dependence and all curves collapse. The increase of $c$ for larger disorder strength $h$ to have the collapse results interesting and probably deserves further research. Our finite-size scaling analysis indicates that the claim of many-body localization suppression due to strong degree of correlations between the disordered potentials could remain valid even at the $L\to\infty$ limit. 

One can infer the dynamical behavior of the time-dependent quantities from the analysis based on the time-independent quantities presented in this SM. For instance, and as already explained in the main text, stronger energy level repulsion produces a deeper repulsion hole, meanwhile absence of level repulsion leads to absence of the repulsion hole. See also the discussion in the introductory part of the dynamical quantities in the main text.


%\bibliography{Correlated.bib} 